# The Influence of Morphology on the Charge Transport in Two-Phase Disordered Organic Systems

Cristiano F. Woellner[1], Leonardo D. Machado[1], Pedro A. S. Autreto[1], José A. Freire[2] and Douglas S. Galvão[1]
[1]Applied Physics Department, State University of Campinas, 13083-970 Campinas, São Paulo, Brazil.
[2]Physics Department, Federal University of Paraná, 81531-990, Curitiba, Paraná, Brazil

## ABSTRACT

In this work we use a three-dimensional Pauli master equation to investigate the charge carrier mobility of a two-phase system, which can mimic donor-acceptor and amorphous-crystalline bulk heterojunctions. Our approach can be separated into two parts: the morphology generation and the charge transport modeling in the generated blend. The morphology part is based on a Monte Carlo simulation of binary mixtures (donor/acceptor). The second part is carried out by numerically solving the steady-state Pauli master equation. By taking the energetic disorder of each phase, their energy offset and domain morphology into consideration, we show that the carrier mobility can have a significant different behavior when compared to a one-phase system. When the energy offset is non-zero, we show that the mobility electric field dependence switches from negative to positive at a threshold field proportional to the energy offset. Additionally, the influence of morphology, through the domain size and the interfacial roughness parameters, on the transport was also investigated.

## INTRODUCTION

High performance conjugated polymers have gained significant attention in recent years due to their low-cost processing and high ductility, fundamental features for applications in flexible electronics. These features provide them with a significant competitive advantages over other technologies based on crystalline semiconductors [1]. However, despite their great potential for applications, weak device performance is still a limiting factor and many fundamental questions regarding the origin of these problems remain unsolved. Thus, a better understanding of charge transport in these materials is necessary in order to overcome these difficulties.

With the advent of solar cells based on bulk heterojunctions (BHJs) [2,3] the need for models that describe the charge transport in such systems increased. In 2005 Watkins *et al.* [4] proposed an effective model based on the dynamical Monte Carlo method to generate the morphologies of binary mixtures, and in 2010, using a proper combination of Pasveer [5] and Watkins approaches, Koster [6] showed for the first time that the mobility in donor-acceptor blends could exhibit a negative electric field dependence.

Despite these advances, all models so far for charge transport in BHJs explicitly assumed that the transport occurs exclusively in one phase (donor or acceptor), independently of the difference in energy between the electronic states of the two phases or on the applied electric field value [7].

## THEORY

In this work we investigated the charge carrier mobility of a two-phase system using an approach that can be separated into two parts: the morphology generation and the charge transport modeling in the generated blend. The morphology part is based on a lattice-gas model of a binary mixture developed by Watkins *et al*. [4]. The system is defined on a regular cubic lattice of $N$ sites and lattice parameter **a**. The phase-1 is constituted by $\alpha N$ sites and phase-2 by $(1-\alpha)N$ sites, where $\alpha$ is the volume ratio. The initial lattice configuration is a random mixture of the constituents with a fixed $\alpha$. In order to simulate a realistic two-phase system, such as donor-acceptor or amorphous-crystalline blends, phase segregation is induced. This is accomplished thorough Monte Carlo simulation by adjusting the interaction energy between the constituents. At every Monte Carlo step a pair of neighboring sites is randomly chosen and the total energy of the system before and after the sites swap their positions is calculated. If the total energy decreases the swap is automatically accepted, otherwise a non-zero probability of acceptance is associated with the exchange.

The phase separation is characterized by two parameters: a characteristic length, known as the domain size, defined by [6,8]:

$$b = \frac{6(1-\alpha)V}{A}, \tag{1}$$

where $V$ is the total volume and $A$ is the interfacial area. The second parameter is associated with the interfacial roughness between the two phases, named the interfacial roughness parameter, $\gamma$. The $\gamma$ values are defined controlling the interaction energy between the sites (black/white, see Figure 1) of which phase. Lower $\gamma$ values imply high roughness values. Figure 1 illustrates two typical cases: a rougher interfacial surface ($\gamma$=0.1) and a smoother one ($\gamma$=0.7). The Monte Carlo simulation is continuously carried out until the desired domain size is reached.

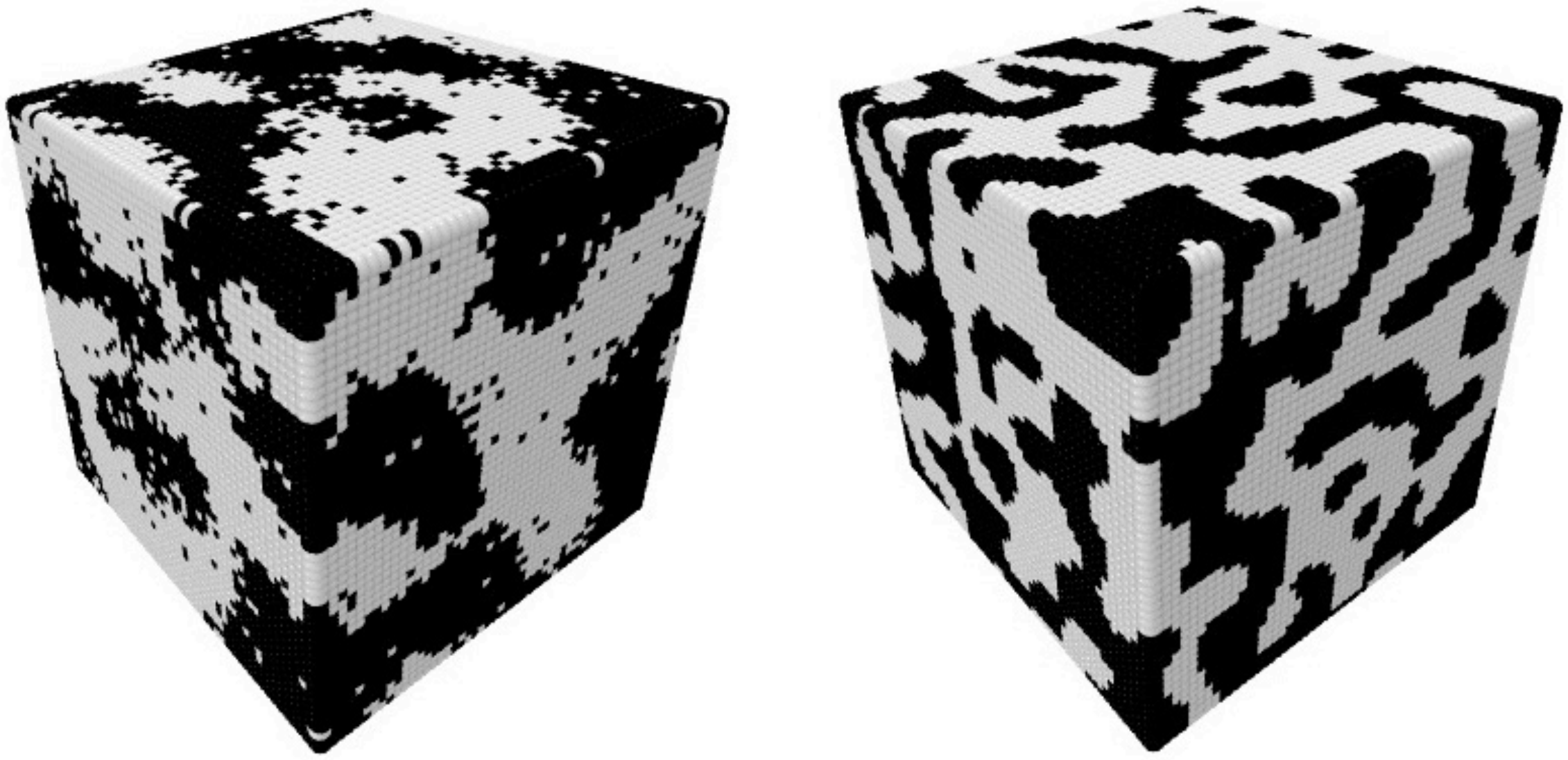

**Figure 1**. Two representative morphologies with an equal amount of constituents (volume ratio $\alpha$=0.5) and the same domain size (b=6 nm). The left case illustrates a rougher interfacial surface ($\gamma$=0.1) and the right one illustrates a smoother surface ($\gamma$=0.7).

Disordered organic materials, due to the large morphological disorder and the weak electronic coupling, have localized electronic states. The energetic distribution of these localized states in a one-phase system is usually assumed to be Gaussian [9]. In the present work we will consider just the electrons. For the two-phase case it can be assumed a bimodal Gaussian density of states, where each phase is characterized by a Gaussian distribution with width $\sigma_1$ (phase-1) and $\sigma_2$ (phase-2). Hereafter, we will assume $\sigma_1 = \sigma_2 = 0.1$ eV.

The charge carrier mobility is calculated by numerically solving the steady-state Pauli master equation:

$$\sum_{j\neq i}[W_{i\to j}P_i(1-P_j)-W_{j\to i}P_j(1-P_i)]=0, \tag{2}$$

where $P_i$ is the probability that site $i$ is occupied by a charge carrier, and $W_{i\to j}$ is the hopping rate from site $i$ to site $j$. The (1-$P_i$) term excludes, in a mean-field approximation, the possibility of double occupancy. The hopping rate is assumed to be of the Miller-Abrahams form and the electric field, $F$, is assumed constant through the system.

We solved Eq. (2) for the occupational probabilities, $P_i$'s, using periodic boundary conditions by an iteration procedure and once the occupational probabilities are obtained, the charge-carrier mobility $\mu$ is calculated from:

$$\mu=\frac{\sum_{i,j\neq i}W_{i\to j}P_i(1-P_j)R^x}{nFV} \tag{3}$$

where $n=\langle P_i\rangle/a^3$ and $V$ is the system volume. The lattice size is $N=100^3$. Averages over a number of different disorder configurations and different morphologies (with fixed domain size, interfacial roughness and volume ratio) were taken until accuracy better than 10% was obtained for $\mu$. Throughout the paper we used fixed values for the carrier density ($n=10^{-5}a^{-3}$), the lattice constant ($a$=1 nm), and the thermal energy ($k_BT$=0.025 eV, corresponding to room temperature).

## DISCUSSION

In figure 2 we show, using a linear scale, the effect of the domain size, $b$, and the interfacial roughness, $\gamma$, on the mobility, normalized to the "$b$=6 nm" and "$\gamma$=0.7" mobility $\mu_0=1.12\text{x}10^{-4}$ cm$^2$/Vs. For $\gamma$ larger than some threshold interfacial roughness parameter (here $\gamma$>= 0.2) we can clearly seen from the Figure that increasing $b$ also enhances the mobility. The domain size $b$ has a direct impact on the channel network of each phase. As it increases the number of direct percolative paths from one electrode to the other, consequently this increases the mobility. But for a rougher interfacial surface, $\gamma < 0.2$, the carrier mobility does not increases monotonically but has a maximum value at a specific domain size value. This means that the optimal value for the carrier mobility is achieved by a right combination of both domain size and interfacial roughness parameters.

## ELECTRIC FIELD DEPENDENCE

We have discussed in the last section the effect of morphology in the zero field regime. In this section we will discuss the influence of the electric field on the mobility for the case presented in the previous section: both phases equally disordered. We will assume here $b$=6 nm and $\alpha$=0.5.

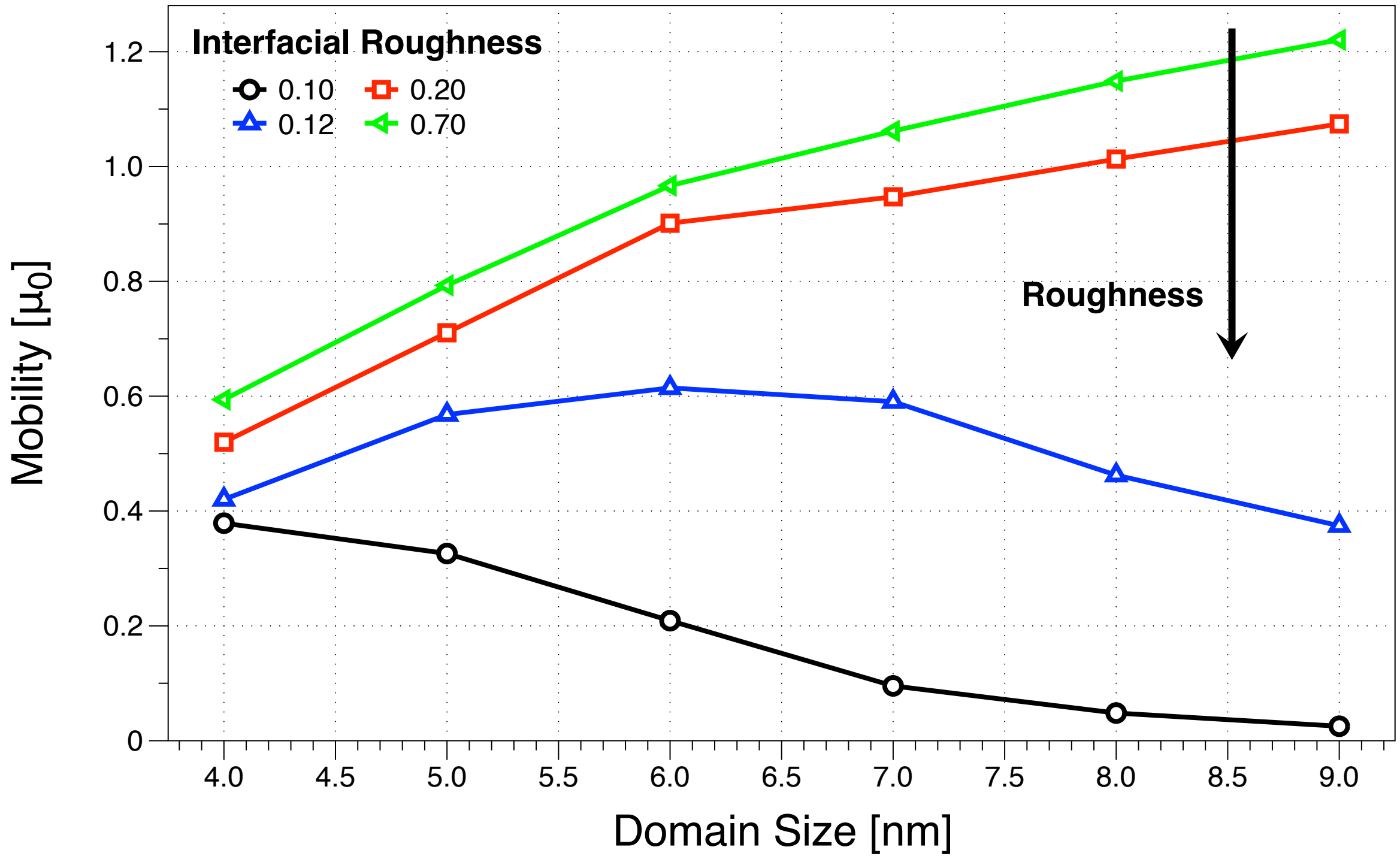

**Figure 2**. Mobility values as a function of the domain size, $b$, for selected values of the interfacial roughness, $\gamma$. For all the investigated cases we used the volume ratio $\alpha$=0.5, and both phases are equally disordered, $\sigma_1 = \sigma_2$=0.1 eV. The mobility is normalized to the "$b$=6 nm" mobility $\mu_0=1.12\text{x}10^{-4}$ $cm^2$/Vs.

In Figure 3 we show, in a semi-log scale, the effect of the electric field, $F$, on the mobility values. We fixed $\sigma_1=\sigma_2$=0.1 eV and considered four different values of $\gamma$. For $E_{offset}$ >0 (here $E_{offset}$=0.1 eV), the mobility has a negative field dependence followed by positive one [10]. The negative field dependence was discussed by Bässler [9] and Koster [6] in the context of a single-phase system. The effect was explained in terms of paths that go against the field that contribute to the mobility at very low fields but cease to contribute (hence decreasing the mobility) as the field increases. Above a certain $F_{min}$ only paths that go mostly along the field contribute and the mobility along these paths increases with increasing $F$. What is displayed in figure 3 is a somewhat similar effect on a two-phase system. For $F$ lower than some threshold $F_{min}$ the carrier is entirely restricted to the less energetic phase-2 at low fields, the observed decrease of $\mu$ below $F_{min}$ is the effect of the field in the transport through the channel network of phase-2. Above $F_{min}$ the field can provide energy for the carrier to hop into the more energetic phase-1 (it is significant that $eF_{min}a$ is of the order of $E_{offset}$), and a number of paths that were forbidden at low fields become available, resulting in a mobility increase. Finally, Figure 3 also shows that the roughness affect is more pronounced for higher electric field values.

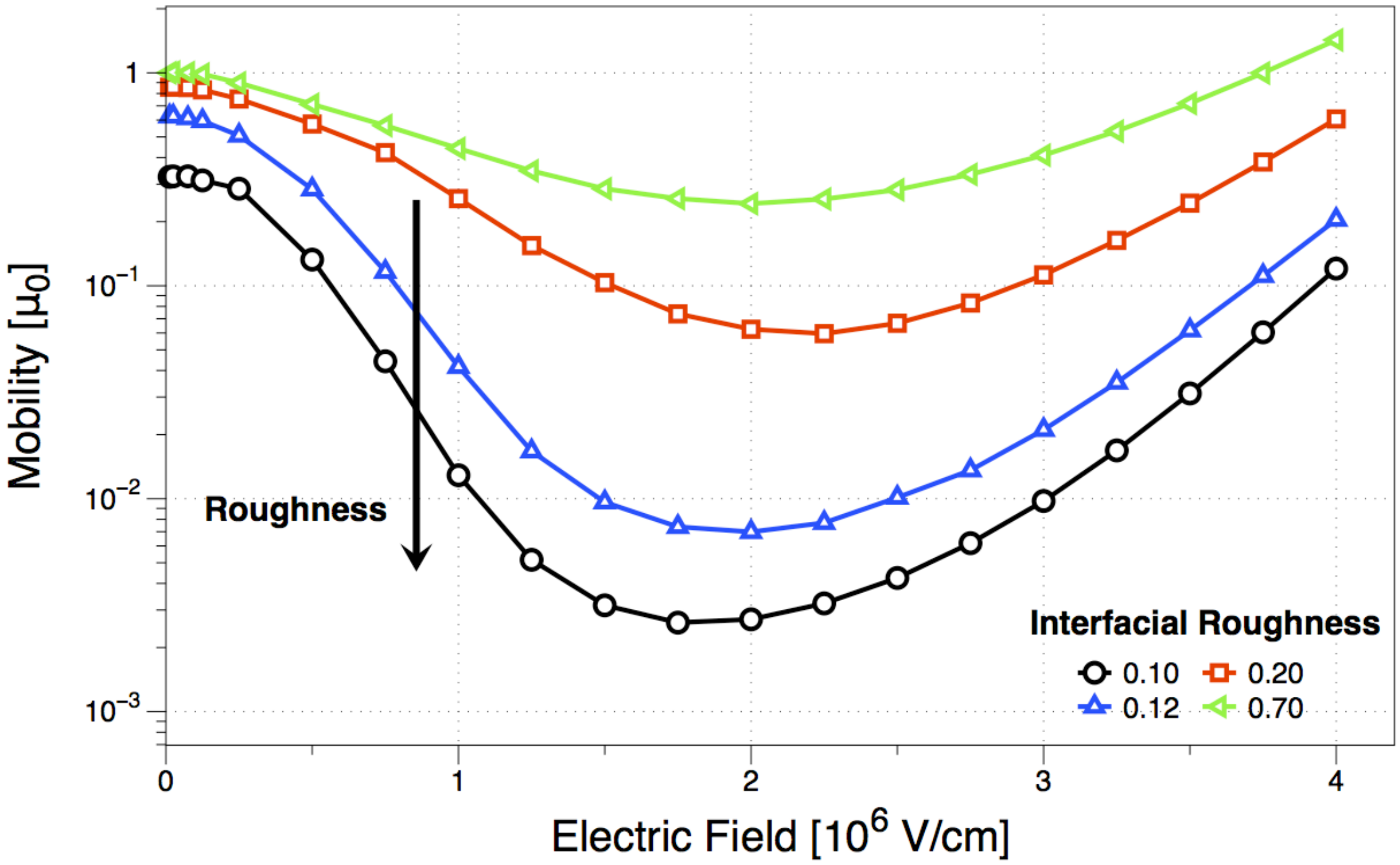

**Figure 3**. Mobility values as a function of the electric field, for selected values of the interfacial roughness $\gamma$. For all cases, the domain size was $b$=6 nm, the volume ratio $\alpha$= 0.5, and both phases are equally disordered, $\sigma_1=\sigma_2=0.1$ eV. The mobility values are normalized to the zero-field "$b$=6 nm" mobility $\mu_0=1.12 \times 10^{-4}$ $cm^2/Vs$. We can clearly see that the roughness strongly affects the mobility values.

## CONCLUSIONS

We have applied a three-dimensional Pauli master equation model with a bimodal Gaussian density of states to investigate the influence of morphology and electric field on the charge carrier mobility of a two-phase system in the low carrier density limit. At low electric fields and the two phases being equally disordered, we showed that the carrier mobility have a completely different behavior depending on the domain size, $b$, and the interfacial roughness, $\gamma$, parameters. For the same domain size, the carrier mobility decreases with increasing the roughness. For each value of the interfacial roughness parameter the carrier mobility has a maximum value at a specific domain size value. This suggests that the optimal value for the carrier mobility can be obtained by a right combination of both parameters. We have also shown that the electric field dependence on the mobility shows a minimum value which is strongly dependent on the interfacial roughness. In practice, these findings provide in an approximate way, for which regime (negative or positive mobility field dependence) a particular system, for instance a donor-acceptor blend will operate at a given electric field.

## ACKNOWLEDGMENTS

This work was supported in part by the Brazilian Agencies CAPES, CNPq and FAPESP. The authors thank the Center for Computational Engineering and Sciences at Unicamp for financial support through the FAPESP/CEPID Grant # 2013/08293-7.